\documentstyle[sprocl,epsfig]{article}
\begin{document}

\hspace*{\fill} KIAS-P99094

\hspace*{\fill} hep-ph/9910325

\vskip 0.2cm

\title{ DETERMINING SUSY PARAMETERS IN CHARGINO PAIR \\
        PRODUCTION IN $e^+e^-$ COLLISIONS} 

\author{ S.Y. Choi \footnote{e-mail: sychoi@kias.re.kr}}

\address{ Korea Institute for Advanced Study, 207-43, Cheongryangri-dong,\\
          Dongdaemun-gu, Seoul 130-012, Korea}


\maketitle\abstracts{ In most supersymmetric theories, 
  charginos $\tilde{\chi}^\pm_{1,2}$ belong to the class of the lightest 
  supersymmetric particles and they are easy to observe at $e^+e^-$ colliders. 
  By measuring the total cross sections and the left--right asymmetries with 
  polarized beams as well as the angular correlations of the decay products 
  of the charginos in $e^+e^-\rightarrow\tilde{\chi}^-_i\tilde{\chi}^+_j\, 
  [i,j=1,2]$, the chargino masses and the gaugino--higgsino mixing angles 
  can be determined very accurately. From these observables the relevant 
  fundamental SUSY parameters can be derived: $M_2$, $|\mu|$, $\cos\Phi_\mu$, 
  and $\tan\beta=v_2/v_1$. The solutions are unique. }

\section{Introduction}

In most supersymmetric theories, charginos $\tilde{\chi}^\pm_{1,2}$, mixtures
of charged gauginos and higgsinos, belong to the class of the lightest 
supersymmetric particles and they are easy to observe at $e^+e^-$ colliders.
Once the charginos are discovered, the priority will be to measure 
the low-energy SUSY parameters independently of theoretical models 
and then check whether the correlations among parameters support 
a given theoretical framework, like SUSY-GUT relations.  
For this purpose, it is very important that the $e^+e^-$ c.m. energy 
can be optimized to cross only very few thresholds at a time, and
also to have beam polarization. Making judicious choices of these 
features, the confusing mixing of many final states with the cascade decays 
can be avoided and analyses restricted to a specific subset of processes 
performed. 

In this contribution, we discuss methods of extracting 
SUSY parameters from the chargino sector. We analyze attempts of ``measuring'' 
the fundamental parameters at a $e^+e^-$ linear collider (LC) by taking
two steps \cite{choi1,choi2}: (I) From the observed quantities such as 
cross sections, polarization asymmetries, and angular correlations we 
determine the phenomenological parameters: the chargino masses and mixings, 
and (II) from the phenomenological parameters we extract the Lagrangian 
parameters: $M_{1,2}$, $\mu$, $\tan\beta$, and $m_{\tilde{\nu}}$.
Each step can suffer from both experimental problems and theoretical
ambiguities. 
An alternative approach for the step II, based only on
the masses of some of the  charginos and neutralinos, has been provided by
Kneur and Moultaka \cite{KM}, and the polarization and spin effects in the
neutralino system has been studied by Moortgat-Pick {\it et al.} \cite{gudi}.
In contrast, many earlier analyses \cite{other} have elaborated on global 
fits.

\section{Determining SUSY parameters in the chargino system}

The spin--1/2 superpartners of the $W$ boson and charged Higgs boson,
$\tilde{W}^\pm$ and $\tilde{H}^\pm$, mix to form chargino mass
eigenstates $\tilde{\chi}^\pm_{1,2}$ through the chargino mass matrix in the
$(\tilde{W}^-,\tilde{H}^-)$ basis
\begin{eqnarray}
{\cal M}_C=\left(\begin{array}{cc}
                M_2                &      \sqrt{2}m_W c_\beta  \\
             \sqrt{2}m_W s_\beta  &             \mu   
                  \end{array}\right)
\label{eq:mass matrix}
\end{eqnarray}
which is given in terms of fundamental parameters: $M_2$, $\mu$, and
$\tan\beta=v_2/v_1$; $s_\beta=\sin\beta$, $c_\beta=\cos\beta$.
In CP--noninvariant theories, the gaugino mass $M_2$ and the 
higgsino mass parameter $\mu$ can be complex. 
However, by reparametrization of the
fields, $M_2$ can be assumed real and positive without loss of
generality so that the only non--trivial invariant phase is 
attributed to $\mu$: $\mu=|\mu|\,{\rm e}^{i\Phi_\mu}$ ($0\leq \Phi_\mu\leq
2\pi$).

The complex, asymmetric chargino mass matrix ${\cal M}_C$ is diagonalized
with two different unitary matrices $U_L$ and $U_R$ acting on the left-- 
and right--chiral $(\tilde{W}^-,\tilde{H}^-)$ states, respectively,
which can be parametrized in terms of the left/right mixing angles
$\phi_{L,R}$, the left/right CP--violating 
phases $\{\beta_L,\beta_R\}$ and two additional phases $\{\gamma_1,
\gamma_2\}$. These phases are not independent but can be expressed 
in terms of the invariant angle $\Phi_\mu$, and all of them  
vanish in CP--invariant theories for  $\Phi_\mu=0$ or $\pi$. 
The mass eigenvalues $m^2_{\tilde{\chi}^\pm_{1,2}}$ and the rotation angles 
$\phi_L$ and $\phi_R$ are uniquely determined by the fundamental SUSY 
parameters $\{\tan\beta, M_2,|\mu|, \cos\Phi_\mu\}$.

Charginos are produced either in diagonal or in mixed pairs in $e^+e^-$ 
collisions. With the second chargino $\tilde{\chi}_2^\pm$ 
expected to be significantly heavier than the first one, 
$\tilde{\chi}_1^\pm$ may be, for some time, the only chargino state 
that can be studied experimentally in detail in the first phase
of a LC.  Keeping in mind this point, we study two cases - (i) only the 
lightest charginos can be pair produced 
without beam polarization in the CP invariant theories and (ii)  
any pairs of chargino states can be produced with sufficient energies and 
beam polarization in the CP noninvariant theories.

\subsection{Diagonal pair production of light charginos in the CP invariant
            theories}

The $\tilde{\chi}^-_1\tilde{\chi}^+_1$ production cross section  in
$e^+e^-$ collisions depends on $m_{\tilde{\chi}^\pm_1}$, $m_{\tilde{\nu}}$ 
and the mixing angles $\phi_{L,R}$ and it increases very sharply near 
threshold, allowing the precise determination of the chargino mass. 

Each of the charginos decay directly to a pair of fermions 
and the (stable) lightest neutralino $\tilde{\chi}_1^0$ through the
exchange of a $W$ boson or scalar partners of fermions. In that case,
two invisible neutralinos in the final state of the process, $ e^+ e^-
\to \tilde{\chi}_1^+ \tilde{\chi}_1^- \to \tilde{\chi}_1^0
\tilde{\chi}_1^0 (f_1 \bar{f}_2) (\bar{f}_3 f_4)$, makes it impossible
to measure directly the chargino production angle $\Theta$ in the
laboratory frame.  Integrating over this angle and also over the
invariant masses of the fermionic systems $(f_1 \bar{f}_2)$ and
$(\bar{f}_3 f_4)$, we can write the fully--correlated distribution
$\Sigma(\theta^*, \phi^*, \bar{\theta}^*, \bar{\phi}^*)$  in terms of 
sixteen independent angular combinations of helicity production amplitudes:
\begin{eqnarray}
\Sigma &=& \Sigma_{\rm unp}+\kappa (\cos\theta^*+\cos\bar{\theta}^*){\cal P} 
        -\kappa^2\left[\cos \theta^* \cos \bar{\theta^*}{\cal Q}\right. 
	 \nonumber\\
 &&\left. +\sin\theta^*\sin\bar{\theta^*}\cos(\phi^*+\bar{\phi^*}){\cal Y}
    \right] 
    + \dots\,,
\label{Sigma}
\end{eqnarray}
in the CP--invariant theory, where the analysis power $\kappa$ contains 
all the complicated dependence on the chargino decay dynamics such as 
neutralino and sfermion masses and their couplings, and
$\theta^*$ is the polar angle of the $f_1 \bar{f}_2$
system in the $\tilde{\chi}^-_1$ rest frame with respect to the
chargino's flight direction in the lab frame, and $\phi^*$  the
azimuthal angle with respect to the production plane; quantities with
a bar refer to the $\tilde{\chi}^+_1$ decay.  

A crucial observation \cite{choi1} is that all explicitly written terms 
in eq.~(\ref{Sigma}) can be extracted and three
$\kappa$-independent physical observables, $\Sigma_{\rm unp}, {\cal
P}^2/{\cal Q}$ and ${\cal P}^2/{\cal Y}$, constructed by kinematical 
projections with $\cos \theta^*, \cos \bar{\theta^*}$ and $\sin \theta^* 
\sin \bar{\theta^*} \cos( \phi^* +\bar{\phi^*})$ fully determined 
by the measurable parameters $E, |\vec{p}|$ (the energy and momentum 
of each of the decay systems $f_i \bar{f_j}$ in the laboratory frame) 
and the chargino mass. As a result, the chargino properties can be 
determined without any strong dependence on the other sectors of the model.
The measurements of the cross section and either of the ratios 
${\cal P}^2/{\cal Q}$ or ${\cal P}^2/{\cal Y}$, interpreted as contour 
lines in the plane $\{\cos 2\phi_L,\cos 2\phi_R\}$, intersect at 
some discrete points to enable us to determine $\cos2\phi_L$ and
$\cos2\phi_R$. An example for the determination of $\cos 2\phi_{L,R}$
is shown in fig.~1 for $m_{\tilde{\chi}^\pm_1}=95$ GeV and 
the ``measured'' observables; $\sigma_{tot}=0.37\,{\rm pb}$, 
${\cal P}^2/{\cal Q}
=-0.24$ and ${\cal P}^2/{\cal Y}= -3.66$ whose contour lines
meet at a single point $\{\cos 2\phi_L=-0.8,\, \cos 2\phi_R=-0.5\}$ with 
$m_{\tilde{\nu}}=250$
GeV \footnote{$m_{\tilde{\nu}}$ can be determined together with
the mixing angles by requiring a consistent solution from the
``measured quantities'' $\sigma_{tot}$, ${\cal P}^2/{\cal Q}$ and ${\cal
P}^2/{\cal Y}$ at several values of the c.m. energy.}.

\vspace{2cm}
\begin{figure}[htb]
\centerline{ \epsfig{figure=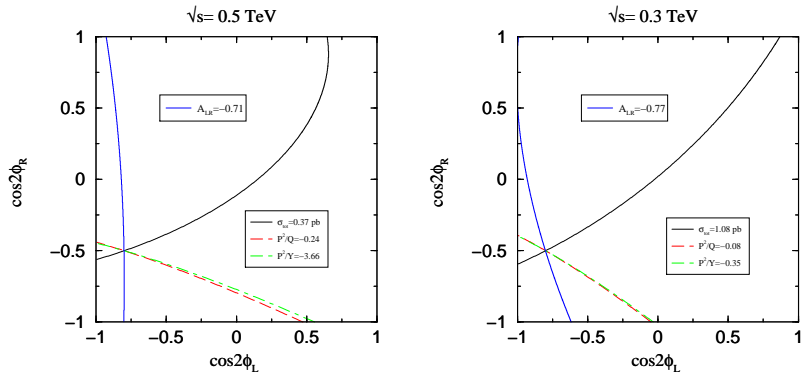, height=3.5cm}}
\caption{Contours for the ``measured values'' of the total cross
  section (solid line), ${\cal P}^2/{\cal Q}$, and ${\cal P}^2/{\cal
  Y}$ (dot-dashed line) for $m_{\tilde{\chi}_1^{\pm}}=95$ GeV
  [\,$m_{\tilde{\nu}} = 250$ GeV\,]. For reference, the contour lines 
  for the "measured" LR asymmetry (solid, almost vertical lines) are
  also superimposed. }
\end{figure}

Let us now discuss the step II by describing briefly how to determine the
SUSY $M_2, \mu$ and $\tan\beta$ from 
$m_{\tilde{\chi}^\pm_1}$, $\cos 2\phi_L$ and $\cos 2\phi_R$ in the 
CP-invariant theory. It is most transparently achieved by introducing 
the two triangular quantities $p=\cot(\phi_R-\phi_L)$ and 
$q=\cot(\phi_R+\phi_L)$.
They are expressed in terms of the measured values $x=\cos 2\phi_L$ and 
$y=\cos 2\phi_R$
up to a discrete ambiguity due to undetermined signs in $p^2-q^2$: 
\begin{eqnarray}
p^2+q^2 =\frac{2(2-x^2-y^2)}{(x-y)^2}, \ \
  pq    =\frac{x+y}{x-y},\ \
p^2-q^2 =\pm\frac{4\sqrt{(1-x^2)(1-y^2)}}{(x-y)^2}\,, 
\end{eqnarray}
which enable us to find {\it at most} two possible solutions
for $\tan\beta$, and then to determine $M_2/\mu=m_W[(p\pm q)s_\beta
-(p\mp q)c_\beta]/\sqrt{2}$, respectively.
The ``measured values'' given in the previous paragraph, for example, 
yield  the results
\cite{choi1}; $[\tan\beta; M_2,\mu] = [1.06; 83\,{\rm GeV}, -59\,{\rm GeV}]$
and $ [3.33; 248\,{\rm GeV}, 123\,{\rm GeV}]$.

To summarize, from the lightest chargino pair production, the 
measurements of the total production cross section and either the
angular correlations among the chargino decay products (${\cal
P}^2/{\cal Q}$, ${\cal P}^2/{\cal Y}$), 
$m_{\tilde{\chi}^\pm_1}$, $\cos 2\phi_L$ and $\cos 2\phi_R$ are determined
unambiguously (Step I). Then the fundamental parameters $\tan\beta$,
$M_2$ and $\mu$ are extracted up to a two-fold ambiguity (Step II).
In addition, if polarized beams are available, the left--right (LR) asymmetry 
$A_{LR}$ can provide an alternative way to extract the mixing angles
(or serve as a consistency check). This is also demonstrated in fig.~1,
where contour lines for the ``measured" values of $A_{LR}$ are 
shown.

\subsection{Diagonal and non--diagonal pair production of charginos in
            the CP noninvariant theories}

If the collider energy is sufficient to produce the two chargino
states in pairs, the above ambiguity can be removed \cite{choi2}. 
The new {\it crucial} ingredient in this case is the knowledge of the
heavier chargino mass. Like for the lighter one,
$m_{\tilde{\chi}^\pm_{2}}$ can be determined very precisely from the
sharp rise of the production cross sections
$\sigma(e^+e^-\rightarrow\tilde{\chi}^-_i\tilde{\chi}^+_j)$.
Moreover, the LR asymmetry 
$A_{LR}$ can provide a powerful way to extract the mixing angles without
any detailed information on the chargino decay dynamics.
We demonstrate this point based on the CP--invariant mSUGRA scenario
{\it RR1}: $\{\tan\beta,m_0,M_{\frac{1}{2}}\}=\{3,100\,{\rm GeV},200\,
{\rm GeV}\}$, giving for the chargino/gaugino and sneutrino masses:
$m_{\tilde{\chi}^\pm_{1,2}}=128/346\,{\rm GeV}$ and
$m_{\tilde{\nu}}=166\,{\rm GeV}$. For the c.m. energy $\sqrt{s}=800$ GeV 
the scenario $RR1$ leads to the following values 
for the cross sections; $\sigma_{tot}(1,1)=0.197$ pb, $\sigma_{tot}(1,2)=0.068$ 
pb, and $\sigma_{tot}(2,2)=0.101$ pb, and the LR asymmetries;
$A_{LR}(1.1)=-0.995$, $A_{LR}(1,2)=-0.911$, and $A_{LR}=-0.668$.
Fig.~2 exhibits the contours in the plane $\{\cos2\phi_L,\cos2\phi_R\}$ plane
for the measured values of three cross sections and three LR asymmetries in 
the diagonal and mixed pair-production processes. All the contours meet at
a common point: $[\cos2\phi_L,\cos2\phi_R]=[0.67,0.85]$. Consequently, 
$\cos2\phi_{L,R}$ along with the chargino masses are uniquely determined.

The measured phenomenological parameters - two chargino masses 
$m_{\tilde{\chi}^\pm_{1,2}}$ and the cosines of two mixing angles 
$\cos 2 \phi_{L,R}$ - enable us to decode the basic
SUSY parameters $\{\tan\beta, M_2,|\mu|,\cos\Phi_\mu\}$: (i) $\tan\beta$ 
is uniquely determined in terms of two chargino masses and two mixing 
angles $\tan\beta=[4-\Delta_{\rm C}(x-y)]^{1/2}/[4+\Delta_{\rm C}(x-y)]^{1/2}$
with $\Delta_{\rm C}=(m^2_{\tilde{\chi}^\pm_2}
-m^2_{\tilde{\chi}^\pm_1})/m^2_W$. (ii) based on  the definition 
$M_2>0$, $M_2$ and $|\mu|$  
reads $M_2/|\mu|=m_W[2\Sigma_{\rm C}\mp\Delta_{\rm C}(x+y)]^{1/2}/2$,
respectively,  
with $\Sigma_{\rm C}=(m^2_{\tilde{\chi}^\pm_2}
+m^2_{\tilde{\chi}^\pm_1}-2m^2_W)/m^2_W$; (iii) $\cos\Phi_\mu$ is given by
\begin{eqnarray}
\cos\Phi_\mu=\frac{\Delta^2_{\rm C}(2-x^2-y^2)-8\Sigma_{\rm C}}{
                   \sqrt{16-\Delta^2_{\rm C}(x-y)^2}
		   \sqrt{4\Sigma^2_{\rm C}-\Delta^2_{\rm C}
		   (x+y)^2}}\,.
\end{eqnarray}
\begin{figure}[htb]
\hspace{3cm}
\epsfig{figure=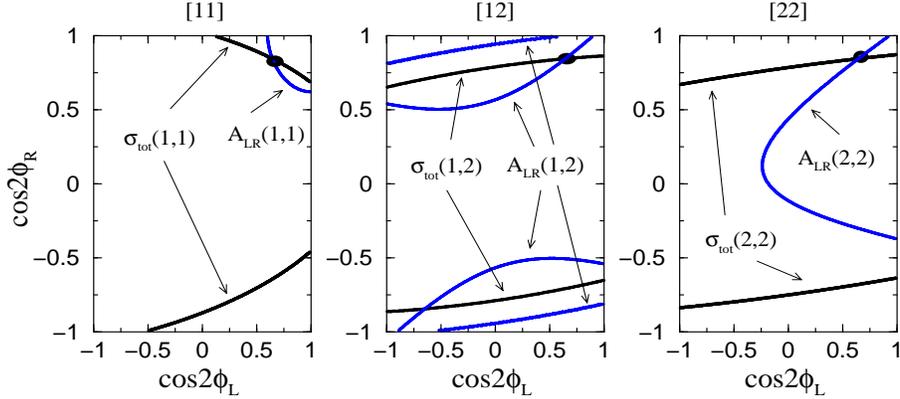, width = 12cm, height= 5.5cm}
\caption{Contours in the $\{\cos2\phi_L,\cos2\phi_R\}$ plane for "measured
    values of $\sigma_{tot}(ij)$ and $A_{LR}(ij)$ $[i,j=1,2]$. The fat dot
    in each figure marks the common crossing point of the contours.}
\label{fig:xrs3}
\end{figure}

As a result, the fundamental SUSY parameters $\{\tan\beta, M_2, |\mu|\, 
\cos\Phi_\mu\}$ in CP--noninvariant theories, 
can be extracted {\it unambiguously} from the observables 
$m_{\tilde{\chi}^\pm_{1,2}}$, $\cos 2\phi_R$, and $\cos 2\phi_L$.
The final ambiguity in $\Phi_\mu \leftrightarrow 2 \pi - \Phi_\mu$ 
may be resolved by measuring observables related to  
the normal polarization of charginos in non--diagonal chargino--pair 
production.
In addition, it is worthwhile to note that the energy distribution 
of the final particles in the decay of the lightest chargino enables 
us to measure the mass of the lightest neutralino as well; this allows 
us to determine the other gaugino mass parameter $M_1$ if it is real. 
Otherwise, 
additional information on the phase of $M_1$ must be derived from
observables involving the heavier neutralinos.

\vskip -0.5cm
\begin{table}[h]
\caption{Mean values and statistical uncertainties of $\{\tan\beta,M_2,
         \mu\}$ obtained with the integrated luminosity of 500 fb$^{-1}$
	 in $RR1$ and $RR2$.}
\vspace{0.4cm}
\begin{center}
\begin{tabular}{|c|c|c|} \hline
{ }           &         $RR1$       &    $RR2$        \\ \hline
$\tan\beta$   &   $3\pm 0.7$        &  ---    \\
$M_2$ [GeV]   &   $152\pm 1.8$      & $150\pm 1.2$    \\
$\mu$ [GeV]   &   $316\pm 0.9$      & $263\pm  0.7$   \\
\hline
\end{tabular}
\end{center}
\end{table}

The strategies presented above are, however, just at the theoretical level. 
The errors in the experimental measurements of physical observables 
need to be estimated to assess fully the physics potential of LC in the
chargino sector. Here, we present just a simple analysis on the expected 
statistical uncertainties with the integrated luminosity of $500$ fb$^{-1}$ 
in $RR1$ and $RR2$ defined
with $\{\tan\beta,m_0,M_{\frac{1}{2}}\}=\{30,160\,{\rm GeV},200\,{\rm GeV}\}$.
Assuming $\Delta m_{\tilde{\chi}^\pm_{1,2}}=0.1$ GeV, the uncertainties in
determining the basic parameters $\tan\beta$, $M_2$ and $\mu$ are present in
Table~1. In both cases, the mass parameters $M_2$ and $\mu$ are determined
with very good precision, but $\tan\beta$ can not be determined unless it
is reasonably small, because of an extremely large error propagations from
$\Delta\cos\beta$ or/and $\Delta\sin\beta$ to $\Delta\tan\beta$ for
a large value of $\tan\beta$, as in $RR2$. 
So, it is very important to find more efficient methods \cite{tanb} to
determine a large $\tan\beta$.

\section{Conclusions}

The chargino masses $m_{\tilde{\chi}^\pm_1}$ and $\cos2\phi_{L,R}$ 
can be extracted from pair production of the lightest chargino pair
in $e^+e^-$ annihilation through the measurement of the production
cross section, the LR asymmetry, and the angular correlations among 
the chargino decay products, even if only the lightest chargino pair 
production is available. With the measured phenomenological parameters, 
$\{\tan\beta, M_2, \mu\}$ can be extracted up to at most a two fold
discrete ambiguity in the CP--invariant theories.

If the c.m. energy is large enough to produce any pairs of charginos,
the measured production cross sections and polarization asymmetries
with beam polarization allow us to determine the chargino masses 
$m_{\tilde{\chi}^\pm_{1,2}}$ and the two mixing angles $\phi_L$ and $\phi_R$ 
very accurately, and to extract the basic SUSY parameters 
$\{\tan\beta,M_2,|\mu|, \cos\Phi_\mu\}$ {\it unambiguously} 
even in the CP--noninvariant theories.

To conclude, the measurement of the processes $e^+e^-\rightarrow
\tilde{\chi}^-_i \tilde{\chi}^+_j$ equipped with polarized beams provides 
a complete analysis of the fundamental SUSY parameters 
$\{\tan\beta, M_2,\mu\}$ in the chargino sector. 

\section*{Acknowledgements}

The author would like to thank his collaborators, A. Djoudi, H. Dreiner, 
J. Kalinowski, H.S. Song and P.M. Zerwas for fruitful collaborations 
and many valuable discussions. 
This work was supported by the Korea Science and Engineering Foundation 
(KOSEF) through the KOSEF-DFG large collaboration project, 
Project No. 96-0702-01-01-2.

\section*{References}

%

%

\end{document}